 \documentclass[final,1p,times]{elsarticle}

\usepackage{amssymb}

\usepackage{hyperref}
\journal{Materials Today Communications}

\begin{document}

\begin{frontmatter}



\title{The New High-entropy Compound RhMnFeCoGe$_4$ with Cubic Non-centrosymmetric B20 Structure}


\author[inst1]{V.~A.~Sidorov}

\affiliation[inst1]{organization={Vereshchagin Institute for High Pressure Physics, RAS},
            city={Troitsk, Moscow},
            postcode={108840}, 
            country={Russia}}

\author[inst1]{V.~N.~Krasnorussky}

\author[inst1]{A.~V.~Bokov}

\author[inst1,inst3]{Z.~N.~Volkova}
\affiliation[inst3]{organization={Mikheev Institute of Metal Physics, Russian Academy of Sciences},
            city={Yekaterinburg},
            postcode={620990}, 
            country={Russia}}
\author[inst1,inst3]{A.~P.~Gerashchenko}

\author[inst1]{N.~M.~Chtchelkatchev}
\author[inst1]{M.~V.~Magnitskaya}
            
\author[inst1]{D.~A.~Salamatin}

\author[inst1]{A.~V.~Semeno}


\author[inst1]{V.~V.~Brazhkin}

\author[inst1]{A.~V.~Tsvyashchenko}

\begin{abstract}
A novel high-entropy compound, RhMnFeCoGe$_4$, with a cubic non-centrosymmetric B20 structure, has been synthesized under conditions of high pressure and temperature.
The electrical transport and magnetic properties of the obtained compound at both ambient and elevated pressures have been investigated.
In addition, nuclear magnetic resonance (NMR) spectra were obtained at 4.2 K and \emph{ab initio} calculations were performed.
The new material exhibits ferromagnetic behavior with a critical temperature of $T_C=146$~K and a spontaneous moment of 2.5 $\mu_B$ per formula unit.
The magnetization data obtained at the critical region yielded the critical temperature and exponents, which were found to be $T_C=146\pm1$~K, $\beta=0.337\pm0.001$, $\gamma=1.121\pm0.001$, and $\delta=4.326\pm0.001$.
The magnetic moments of Mn and Co were determined from NMR spectra to be 2.2~$\mu_B$ and 0.5~$\mu_B$, respectively.
\emph{Ab initio} calculations yielded reasonable values for the lattice parameter and the magnetic moments of all constituents. 
The density of states and band structure are determined for both paramagnetic and ferromagnetic states.
Lattice compression results in the increase in the $T_C$.

\end{abstract}


\begin{highlights}
\item A novel high-entropy compound, RhMnFeCoGe$_4$, with a B20 structure has been synthesized.
\item Pressure dependence of Curie temperature of RhMnFeCoGe$_4$ was measured.
\item Critical behavior analysis led to determine $\beta=0.337$, $\gamma=1.12$, and $\delta=4.32$ of RhMnFeCoGe$_4$ compound.
\item Lattice compression results in the increase in the critical temperature.
\end{highlights}


\end{frontmatter}


\section{Introduction}
\label{sec:introducrion}

Binary transition metal compounds, designated TMX (where TM = Mn, Fe, Co, Rh, and X = Si, Ge), exhibit a cubic noncentrosymmetric structure of type B20. 
The unit cell of the B20 structure is constituted by two mutually penetrating tetrahedrons, formed from four transition metal atoms (TM) and four silicon or germanium atoms (X). 
These compounds are currently the subject of considerable interest as a result of their diverse and intriguing properties. 
These include the formation of skyrmion lattices in MnSi and FeGe \cite{adams2011longrange, tonomura2012realspace, yu2011near} under an external magnetic field, the occurrence of quantum phase transitions at high pressure in MnSi, FeGe and MnGe \cite{pfleiderer1997magnetic, pedrazzini2007metallic, deutsch2014twostep,martin2016magnetovolume}, the existence of massless electronic excited states, known as Weyl fermions, in RhSi and CoSi \cite{tang2017multiple}, and the coexistence of weak ferromagnetism and superconductivity in RhGe\cite{tsvyashchenko2016superconductivity}.

A number of these properties and several others are associated with the non-centrosymmetric structure of the B20 compounds, which is of particular significance with regard to chirality.
Previously, a number of systems comprising intermediate compounds between different binary systems, such as Fe$_{1-x}$Co$_x$Si or Mn$_{1-x}$Co$_x$Ge, have been the subject of investigation. 
Similarly, these compounds were found to possess intriguing properties. 
Despite the fact that FeSi and CoSi lack long-range magnetic order, such an order has been observed in the Fe$_{1-x}$Co$_x$Si alloys \cite{beille1981helimagnetic}.
In the Fe$_{1-x}$Co$_x$Ge system, a change in the sign of chirality was observed at a specific critical composition ($x_c\approx0.6$) \cite{grigoriev2014flip}. 

In the majority of previous studies of mixed systems, the substitutional doping of a tetrahedral sub-lattice comprising one transition metal with another transition metal has been the focus of investigation. 
In this approach, the introduction of a 3\emph{d} metal (Mn, Fe, Co) substituted by a 4\emph{d} metal (Rh) gives rise to new features, which can be attributed to the differing electronic structures of the 3\emph{d} and 4\emph{d} elements.
The distinctive characteristics of the B20 structure suggest that it may serve as a promising platform for the synthesis of a new class of compounds, namely high-entropy compounds, which are currently the subject of considerable research interest. 
These compounds form an ordered crystal lattice, frequently cubic in structure, in which atoms of various transition metals are randomly arranged at crystal sites.
In some cases, the high-entropy materials have improved properties 
compared to the compounds on which they are based on. These improvements include 
superior mechanical properties,
robustness the superconductivity under high pressure \cite{HEA_Guo_190GPa, HEA_SC_SciRep_Kasem2022}, excellent irradiation tolerance \cite{HEA_JALCOM2023, HEA_irradiation_ActaMat2023}, and others.
The absence of B20 magnetics with high $T_{\mathrm{C}}$ and small spiral period / size of skyrmion
limit their applications. Obtaining the high-entropy compounds with B20 structure could
solve these problems due to the cocktail and lattice distortion effects. 
Recent researches have shown that lattice distortion and mixing entropy of transition metals with Si
atoms are responsible for the increased $T_{\mathrm{C}}$ in the Co$_{1+x}$Si$_{1-x}$ and Mn$_{1-x}$Rh$_x$Si 
compounds with small $x$. 
Additionally, the formation of high-entropy compounds with B20 structure, including (FeMnCo)SiGe, (FeCrCo)SiGe, (FeCrMn)SiGe, (CrMnFeCo)Si and CrMnFeCoNiSi was corroborated through experimental analysis and first-principles calculations.
It was proposed that these compounds exhibit enhanced functional characteristics \cite{B20_HEA_Tang2023, B20_HEA_APLmat_2022}.

In the cubic lattice of B20, four different transition metals (Mn, Fe, Co and Rh) can be introduced into one tetrahedral sub-lattice. The successful synthesis of binary compounds for X = Ge and TM = Mn, Co, Rh in the B20 structure is only possible at high pressure. The objective of the present study was to synthesise the high-entropy compound RhMnFeCoGe$_4$ at high pressure, to examine its magnetic and electronic transport properties, and to contrast the obtained properties with those predicted by first-principles calculations.

\section{Methods}

\subsection{Experimental}

The synthesis experiments were conducted under an applied pressure of 8~GPa, generated using a toroidal camera and a cell comprising monocrystalline NaCl. 
The synthesis of the mixture of pure elements (Mn, Fe, Co, Rh, Ge) was achieved through the application of an electric current and subsequent melting of the mixture under pressure. 

X-ray powder diffraction (XRD) analysis was performed on the synthesized products with the assistance of a Huber G670 diffractometer at room temperature. 

The magnetization of the resulting compound was determined via the vibrating sample magnetometer (VSM) option of the PPMS-9 (Quantum Design) instrument.
The magnetization data were obtained over the temperature ($T$) range of 2 to 260~K and in magnetic fields ($\mu_0H$) up to 9 T.

AC transport properties were measured using a synchronous amplifier (SR830, Stanford Research). 
AC magnetic susceptibility measurements were performed at high hydrostatic pressures up to 6~GPa and low temperatures with the help of a miniature autonomous toroidal camera \cite{petrova2005highpressure}.

The $^{55}$Mn nuclear magnetic resonance (NMR) spectra were obtained with AVANCE III Bruker spectrometer in zero magnetic field, $H = 0$, by measuring an intensity of the solid spin-echo signals acquired at equidistant operating frequencies with the two-pulse sequence $\pi/2 - t - \pi/2 - t -$  echo. 
The width of a $\pi/2$ rf-pulses (radio frequency pulses) does not exceed 1~$\mu$s and the time interval, $t = 10\, \mu$s.

\subsection{Calculations}

\emph{Ab initio} DFT calculations of RhMnFeCoGe$_4$ in nonmagnetic and magnetic states were conducted using the PAW pseudopotential method \cite{kresse1999fromultrasoft} as implemented in the VASP package \cite{kresse1996efficient}, with the PBE–GGA version \cite{perdew1996generalized} of the exchange-correlation potential.
The calculations were converged with a plane-wave cutoff of 340~eV and a reciprocal-space resolution of $\sim0.2$\,\AA$^{-1}$ for Monkhorst-Pack \textbf{k}-point grids. 
The structural relaxation was continued down to forces less than 6~meV/atom.

\section{Results and discussion}

\subsection{XRD results}
\begin{figure}
\centering
\includegraphics[width=1.0\columnwidth]{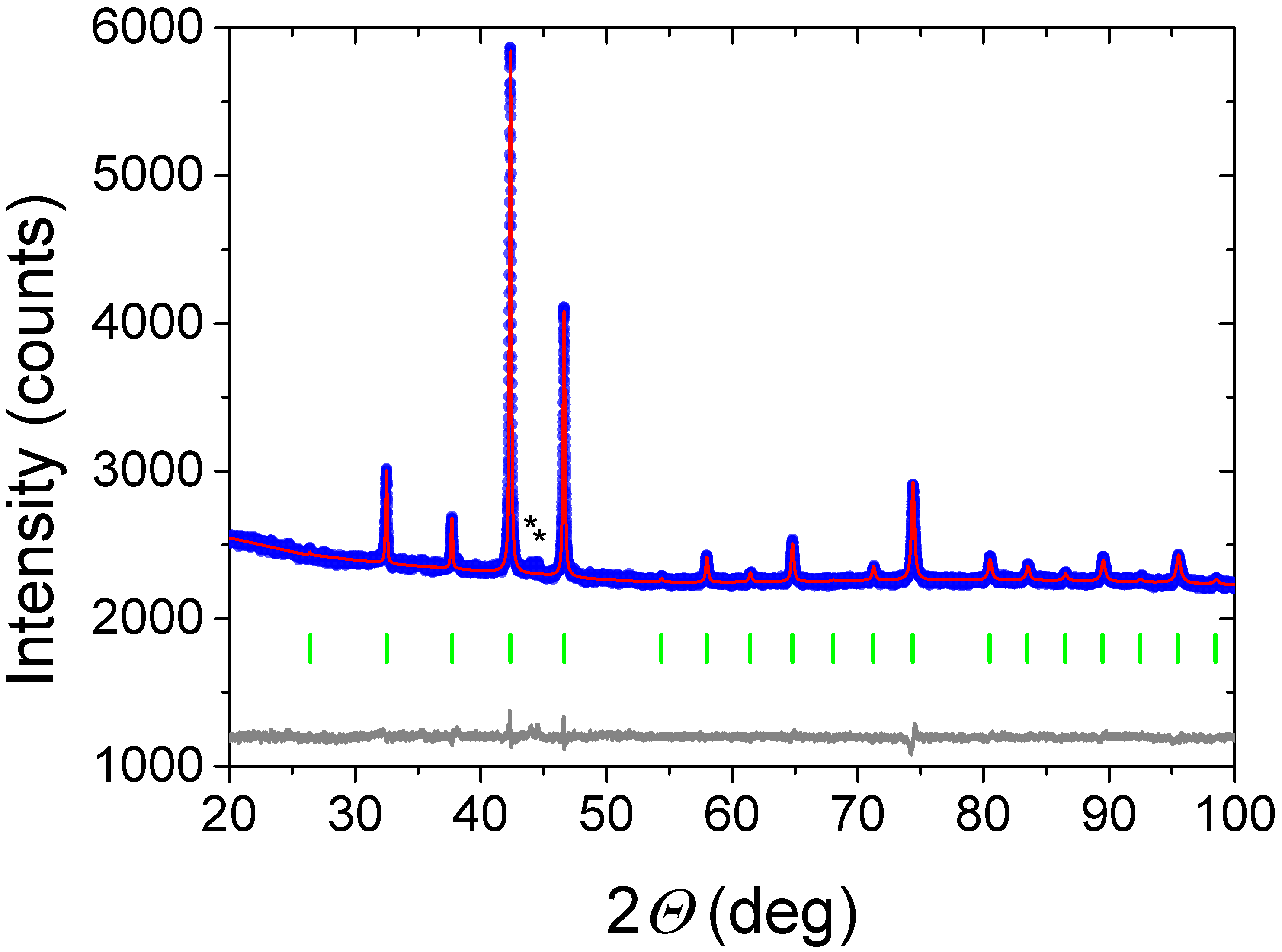}
\caption{The XRD pattern of the RhMnFeCoGe$_4$ compound. The blue dots represent the experimental data, the red line -  fitting by Rietveld method, the gray line depicts the difference between the experiment and fitting, and the green ticks indicate the Bragg positions for the B20 structure.
Asterisks denote the most intense peaks from the Fe$_{1.67}$Ge impurity.}
\label{xrd}
\end{figure}

XRD analysis demonstrated that the RhMnFeCoGe$_4$ samples are single-phase and exhibit a cubic B20 structure in which Rh, Mn, Fe, Co atoms are situated in the one lattice site 4$a$ $x = 0.3842(5)$ with equal probabilities, while Ge atoms are in the different lattice site with $x = 0.0907(5)$. 
The lattice constant of the unit cell is $a = 4.7675(4)$~\AA, which is about 0.01~\AA~larger
in comparison with the average lattice constant for the corresponding germanides. 
On the XRD pattern of the compound (see Fig. \ref{xrd}) there are visible peaks which 
may be attributed to the few wt. \% of the impurity phase of the Fe$_{1.67}$Ge compound 
(usually also denoted as $\beta$-phase in Fe-Ge system with hexagonal unit cell of 
Co$_{1.75}$Ge/Ni$_2$In-type (or B8$_2$ type) crystal structure \cite{Fe-Ge_system_Richardson1967, Fe167Ge_Japan1961, Fe167Ge_Budzynski2015}).

\subsection{Magnetic susceptibility}\label{Sect_suscept}

\begin{figure}
\centering
\includegraphics[width=1.0\columnwidth]{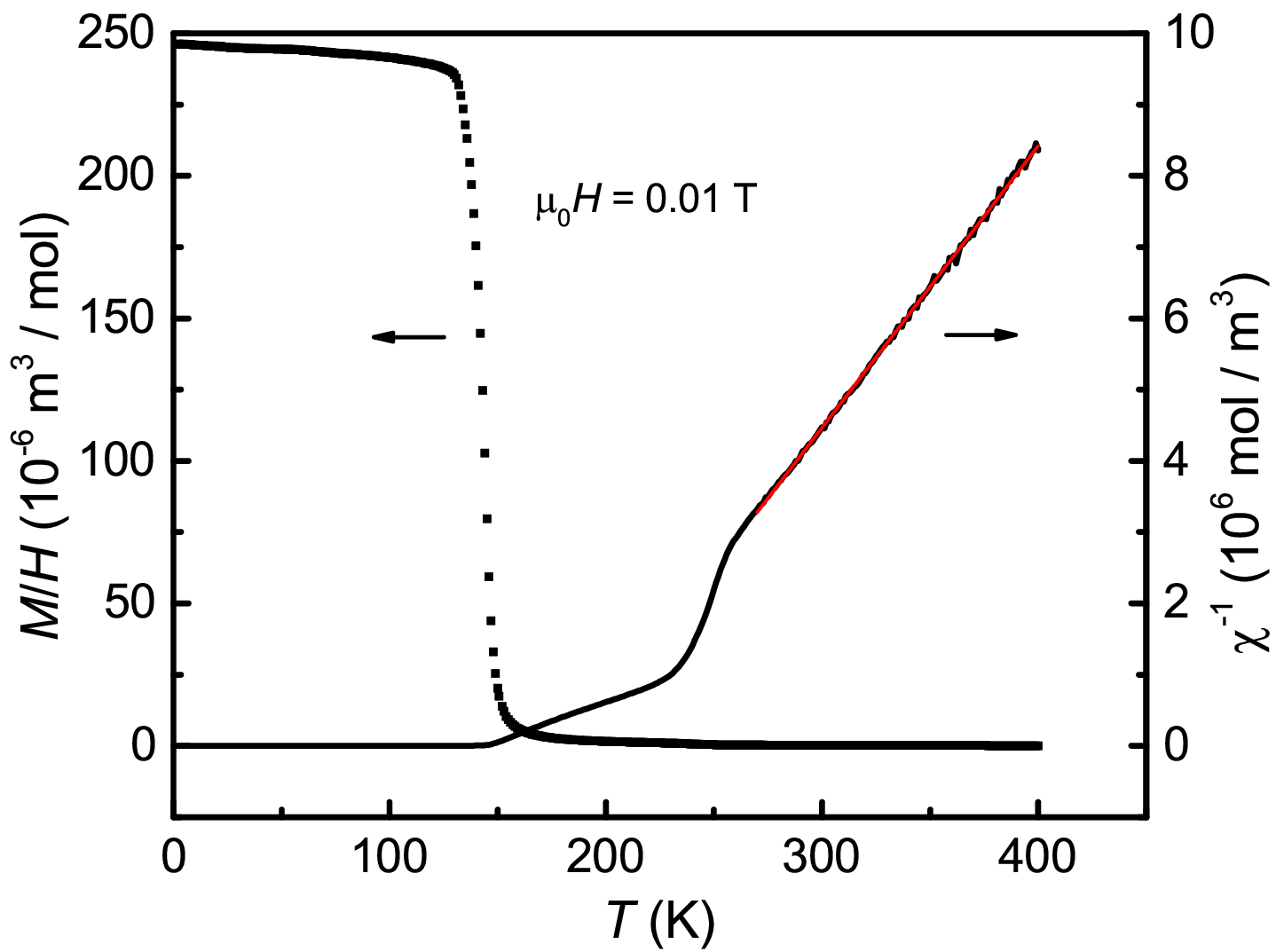}
\caption{The temperature dependence of the magnetic susceptibility $M/(\mu_0H)$ of the RhMnFeCoGe$_4$ compound (left scale), measured in a field of $\mu_0H=0.01$~T, and inverse magnetic susceptibility $\chi^{-1}=(\mu_0H)/M$ as a function of temperature (right scale). Red line - is a Curie-Weiss fit of $\chi^{-1}$ in a linear high-T region with $\mu_{eff}=4.0$ $\mu_B/f.u.$ and Curie-Weiss temperature $\Theta=186$~K.}
\label{hi_inv}
\end{figure}

The results of magnetic susceptibility measurements indicate that, upon cooling below the critical temperature $T_C$, the material undergoes a transition to a magnetically ordered state.
The temperature dependence of the moment is indicative of ferromagnetic ordering.
The temperature dependence of the magnetic susceptibility of RhMnFeCoGe$_4$ was obtained as a ratio of magnetization to an applied magnetic field, $M/(\mu_0H)$, with $\mu_0H=0.01$~T, and is presented in Fig. \ref{hi_inv}.
Additionally, the inverse susceptibility, $\chi^{-1}=(\mu_0H)/M$, is presented on the right scale in Fig. \ref{hi_inv}.
At elevated temperatures, the $\chi^{-1}(T)$ data were analyzed using the Curie-Weiss law $\chi=C/(T-\Theta)$, where $C$ is a Curie-Weiss constant, $\Theta$ is a Curie-Weiss temperature.
The fitted line is illustrated in Fig.\ref{hi_inv} by a red line with parameters $C=2.51\cdot10^{-5}$~m$^3\cdot$K/mol and $\Theta=186$~K.
The effective moment can be obtained from the constant $C$ as $p_{eff}=\sqrt{3k_BC/N_A}$, where $k_B$ is the Boltzmann constant, $N_A$ is the Avogadro number.
Consequently, $p_{eff}=4.0\,\mu_B$/f.u.
A notable departure from the linear Curie-Weiss law of $\chi^{-1}$ dependence is evident within the temperature range from 150 to 250 K.
This departure is related to the emergence of short-range order on cooling within this temperature range.
As demonstrated in Sect. \ref{Sect_magn}, a noticeable spontaneous magnetization arises at 250~K.


\subsection{Magnetization}\label{Sect_magn}

\begin{figure}
\centering
\includegraphics[width=1.0\columnwidth]{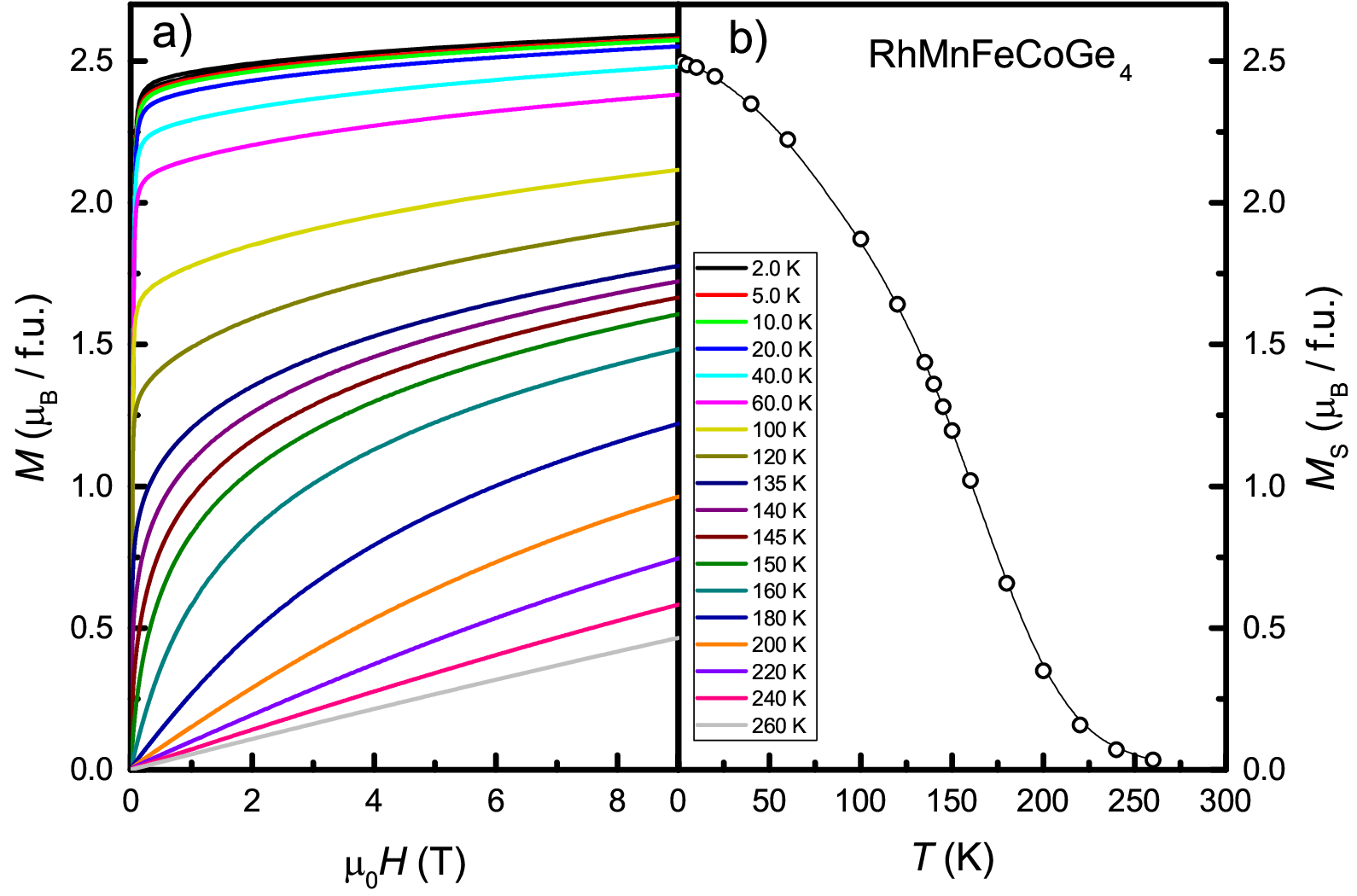}
\caption{Isotherms of magnetization of the RhMnFeCoGe$_4$ compound for temperatures in the range 2-260~K (panel a), spontaneous magnetization $M_S$, obtained at $\mu_0H=9$~T as a function of temperature (panel b).}
\label{m_from_H}
\end{figure}

The results of the magnetization measurements of the RhMnFeCoGe$_4$ compound at temperatures in the range of 2-260~K are presented in Fig. \ref{m_from_H}a.
At temperatures below 120~K, it can be observed that the magnetic moment rapidly approaches saturation in a magnetic field exceeding 0.2~T.
However, at elevated temperatures, the saturation is difficult to achieve due to the weakness of the magnetic field in comparison to thermal excitation.
The temperature dependence of spontaneous magnetization $M_S$ is presented in Fig.\ref{m_from_H}b, obtained by intersecting the tangent line to the magnetization curve at $\mu_0H=9$~T with the ordinate axis. 
The spontaneous magnetization data indicate the presence of a finite magnetization at temperatures below 250~K.
As previously discussed in Sect.\ref{Sect_suscept}, as the temperature is reduced, the short-range order phase emerges before the ordering state becomes established at $T_C\sim150$~K.
As illustrated in Fig. \ref{m_from_H}b, the spontaneous magnetic moment at $T=2$~K is 2.5~$\mu_B$ per formula unit (f.u.).

\begin{figure}
\centering
\includegraphics[width=1.0\columnwidth]{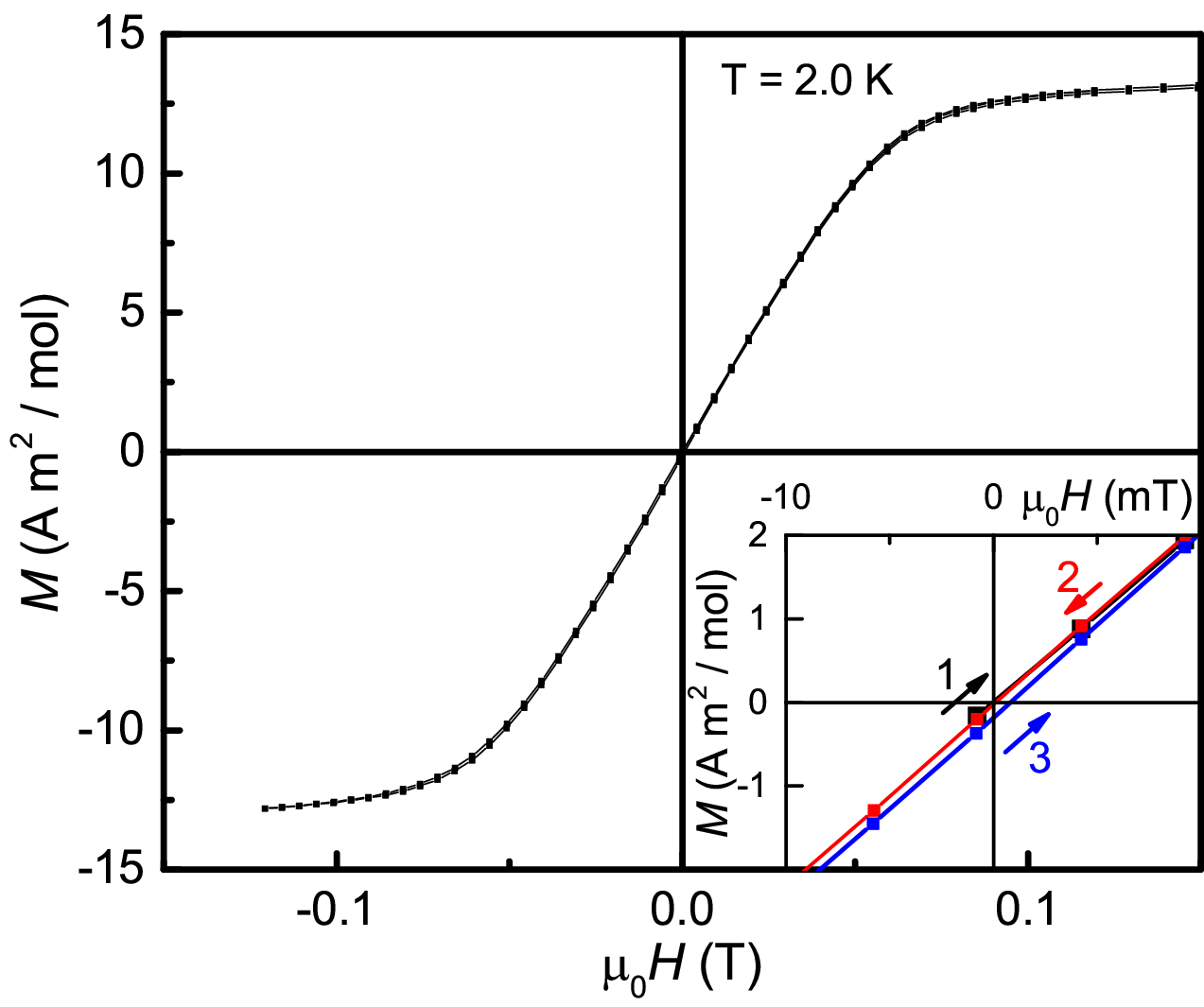}
\caption{The magnetization of the RhMnFeCoGe$_4$ compound, obtained at $T=2.0$~K. Inset - an expanded region in proximity of $\mu_0H=0$. The numbered arrows illustrate the variation in $\mu_0H$, which is represented by the following sequence: 1) an initial increase up to 9~T (black line), 2) a decrease in the field (red), and 3) a subsequent increase (blue).}
\label{hysteresis}
\end{figure}

The magnetization curve of the RhMnFeCoGe$_4$ compound did not exhibit a discernible hysteresis phenomenon.
Figure \ref{hysteresis} illustrates the field dependence of magnetization at a temperature of 2.0~K for an initial increase in field strength up to 9~T, followed by a decrease to -0.12 T and subsequently an increase.
The expanded region in the proximity of zero field is illustrated in the inset.
The arrows illustrate the alteration in field direction for the corresponding colored magnetization line.
Therefore, the coercive force is no greater than 1~mT at $T=2.0$~K, and the remnant magnetization is no more than 0.2~Am$^2$/mol.

\begin{figure}
\centering
\includegraphics[width=1.0\columnwidth]{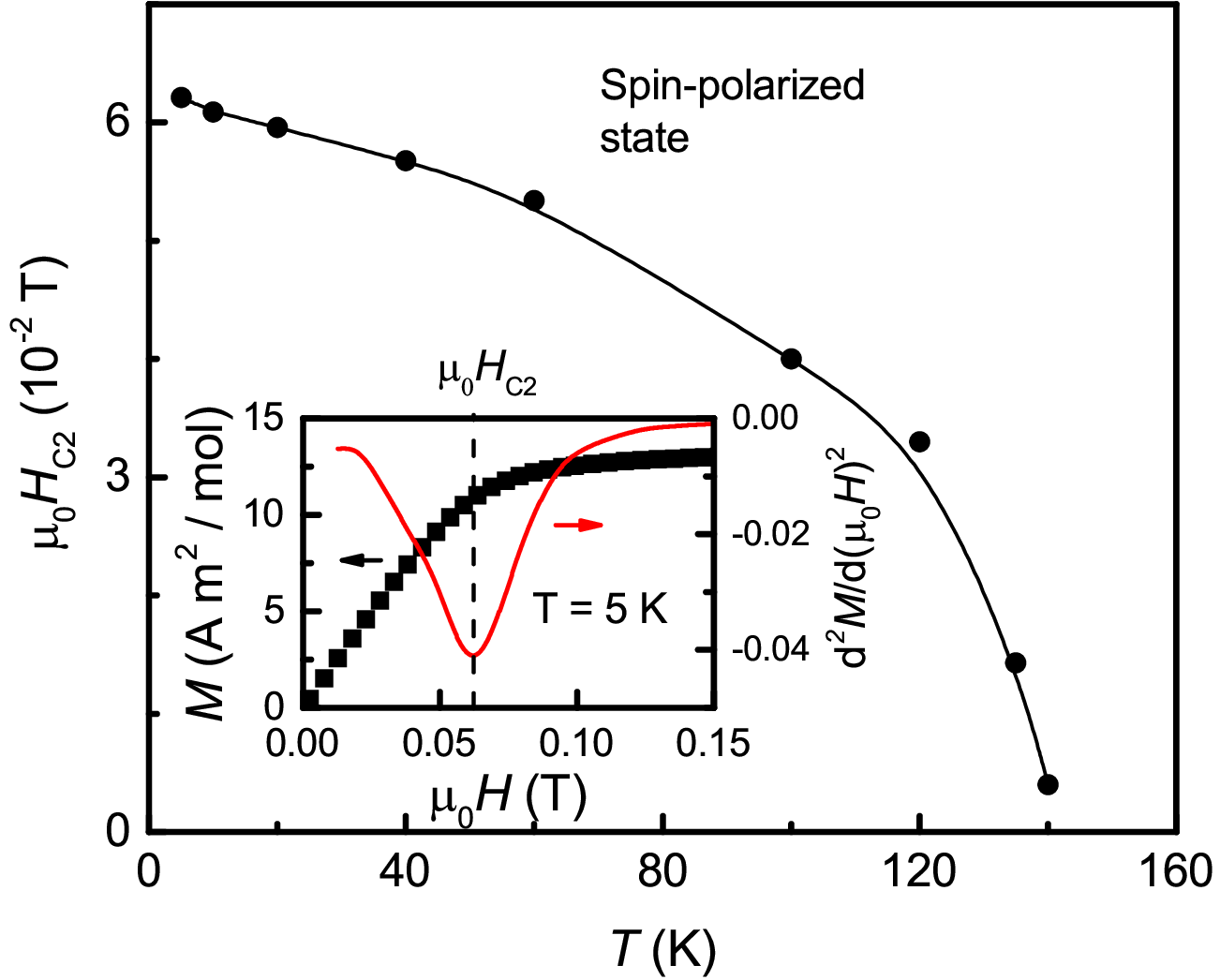}
\caption{The magnetic phase diagram of the RhMnFeCoGe$_4$ compound. The points were identified as local minimums of the second derivative of the magnetization curve, the line is a guide for the eye. Inset demonstrate $M(\mu_0H)$ (points) and $d^2M/d(\mu_0H)^2$ (line) dependencies as a function of magnetic field $\mu_0H$ at $T=5$~K. The position of the $\mu_0H_{C2}$ is indicated by a vertical dashed line.}
\label{phase_diagram}
\end{figure}

In the ordered state, the magnetization curve exhibits a single distinctive feature, manifested as a kink at $\mu_0H \approx 60$~mT at 5 K (inset of Fig.\ref{phase_diagram}). 
The magnetization feature can be readily identified as the minimum of the second derivative of magnetization. 
The phase diagram can be derived from the magnetization curves displayed in Fig.\ref{m_from_H}.
The resulting magnetic diagram is shown in Fig. \ref{phase_diagram}.
This transition line is believed to differentiate between a spin-disordered state and a spin-polarized state, in analogy with the $\mu_0H_{C2}$ field observed in MnSi\cite{narozhnyi2013specific}. 
It is noteworthy that the value of $\mu_0H_{C2}$ at low temperatures for RhMnFeCoGe$_4$ (0.062~T) is approximately one order of magnitude less than that observed in MnSi (0.6~T)\cite{narozhnyi2013specific}.

In the case of weak fields below the critical field value of $H_{C2}$, the magnetization curve of RhMnFeCoGe$_4$ displays a linear increase in magnetization, similar to that observed in MnSi\cite{bloch1975high, narozhnyi2013specific} and FeGe\cite{ditusa2014magnetic}, which possess the same B20 structure.
Upon further augmentation of the magnetic field intensity, a transition to a spin-polarized state is observed in these compounds.
It is therefore reasonable to hypothesize that a comparable helical magnetic structure may emerge in RhMnFeCoGe$_4$.
Nevertheless, to address this question definitively, further studies employing small angle neutron diffraction are necessary.

\subsection{Critical behavior analysis}

\begin{figure}
\centering
\includegraphics[width=1.0\columnwidth]{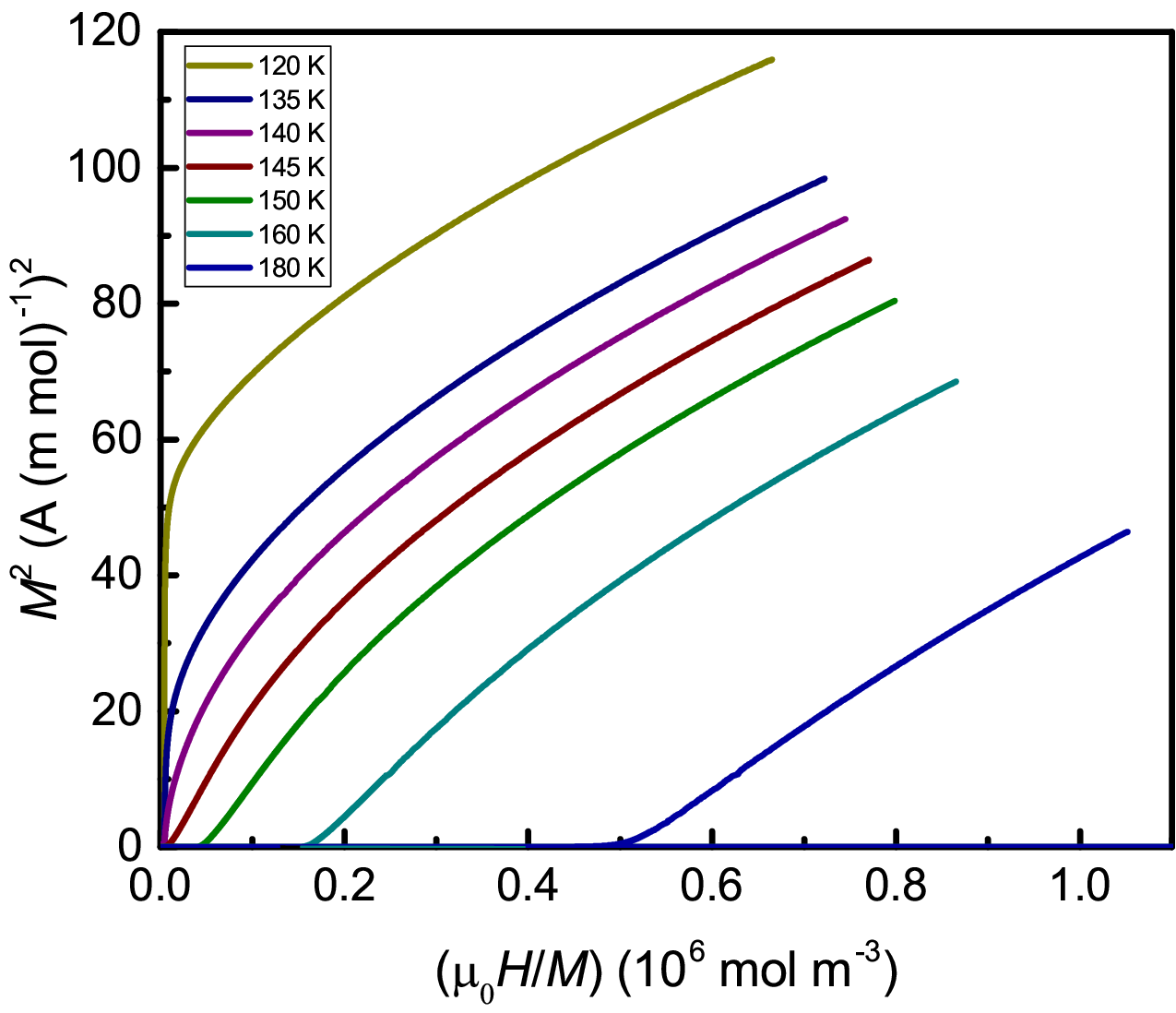}
\caption{$M^2$ vesus $H/M$ (Arrott plot) for RhMnFeCoGe$_4$ compound at various temperatures.}
\label{Arrott}
\end{figure}

The critical transition temperature, $T_C$, of a ferromagnetic material can be determined from an Arrott plot \cite{arrott1957criterion}.
According to this approach,  isothermal magnetization curves  for ferromagnetic materials, represented in the $M^2$ versus $H/M$ coordinates, should exhibit parallelism with one another in magnetic fields of sufficient magnitude, that is, fields exceeding the threshold for magnetic domain reorientation.
The isothermal magnetization curve at some $T=T_C$ in these coordinates at high fields is a straight line crossing the origin.
The Arrott plot for the RhMnFeCoGe$_4$ compound is shown in Fig.\ref{Arrott} at various temperatures in the range from 120 K to 180~K.
The $M^2$ data set does not demonstrate a linear dependence; rather, it exhibits convex curvature.
The convex curvature of $M^2$ curves is a distinctive trait of weak ferromagnets, such as MnSi \cite{bloch1975high} and Fe$_{1-x}$Co$_x$Si \cite{shimizu1990effect}.

\begin{figure}
\centering
\includegraphics[width=1.0\columnwidth]{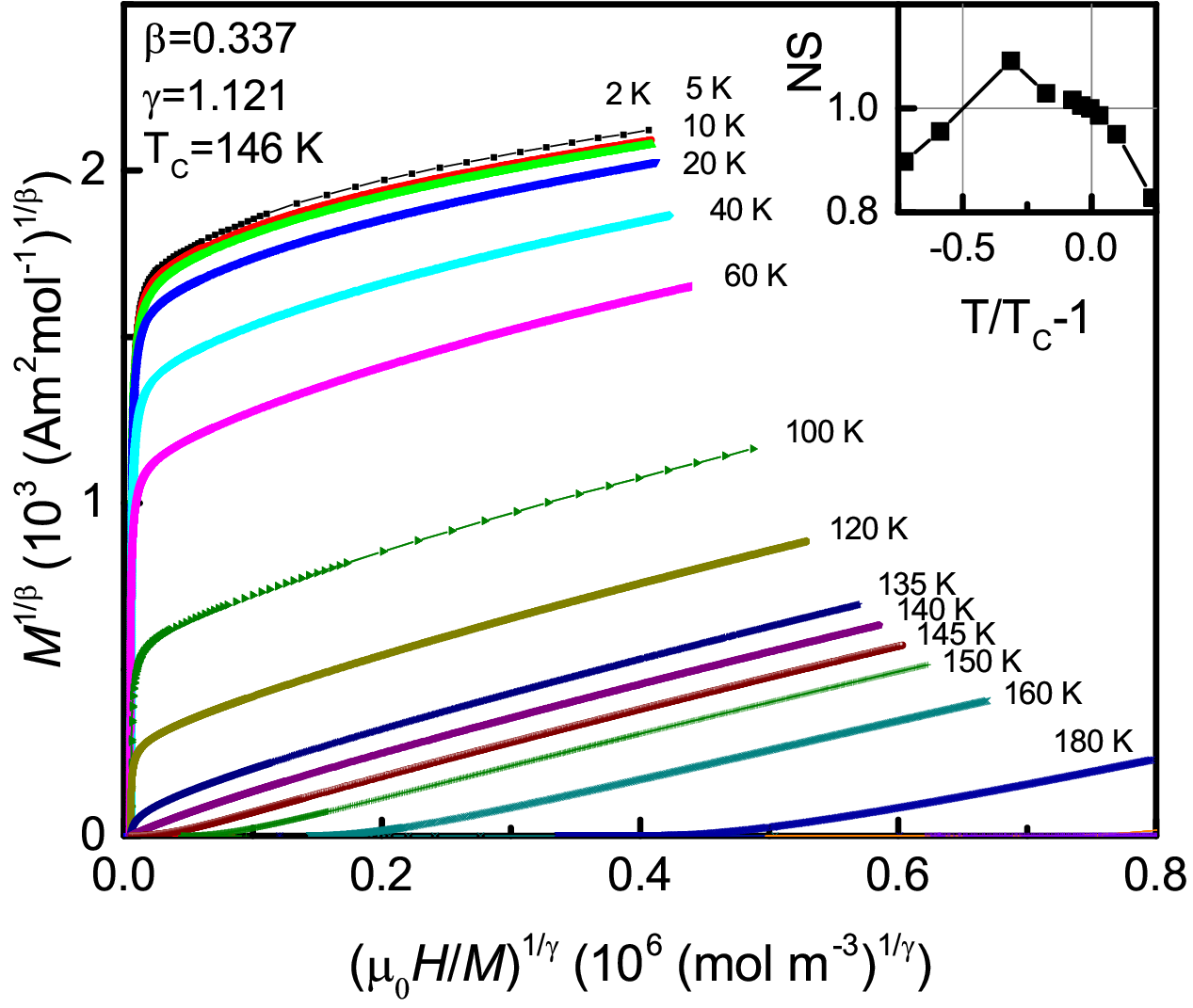}
\caption{A modified Arrott plot for RhMnFeCoGe$_4$ compound with $\beta=0.337$ and $\gamma=1.12$. The inset shows the normalized slope (NS) of the $M^{1/\beta}(\mu_0H/M)^{1/\gamma}$ curves at $\mu_0H=9$~T in the vicinity of critical temperature, $T_C= 146$~K.}
\label{modArrott}
\end{figure}

In order to gain insight into the nature of the phase transition observed at $T_C\approx$140~K, a critical behavior analysis has been conducted. 
The critical behavior was investigated through an analysis of the magnetization curve using a modified Arrott plot.
A modified Arrott plot in  $M^{1/\beta}$ from $(\mu_0H/M)^{1/\gamma}$ coordinates with $\beta=0.34$ and $\gamma=1.12$ is presented in Fig.\ref{modArrott}.
The critical coefficients, $\beta$ and $\gamma$, respectively describe the temperature behavior of the order parameter - spontaneous magnetization $M_S$ and magnetic susceptibility $\chi_0$  in the vicinity of $T_C$, according to the following equations \cite{kaul1985static}: 
\begin{equation}\label{eq_ms}
    M_S(0,T)\sim (-\epsilon)^{\beta}, \, \epsilon<0
\end{equation}
and 
\begin{equation}\label{eq_hi}
    \chi_0(0,T) \sim \epsilon^{-\gamma}, \, \epsilon>0
\end{equation}
where $\epsilon=(T-T_C)/T_C$ is a reduced temperature.
Furthermore, the critical temperature, $T_C=146$~K, was determined through an iterative procedure aimed at identifying the critical coefficients.

The methodology for deriving the critical coefficients $\beta$ and $\gamma$ is outlined in the Refs. \cite{chattopadhyay2009magnetic, krasnorussky2024study}.
The following is a brief overview of the procedure.
From the Arrott plot, the spontaneous moment, $M_S(T)$, and the inverse susceptibility values, represented by the variable, $\chi^{-1}(T)$, were obtained as the intercept of a tangent line to each $M^2(\mu_0H/M)$ curve at $\mu_0H=9$~T with the ordinate or abscissa axes, respectively. 
In the subsequent step of the procedure, the degree indexes $\beta$ and $\gamma$ in the equations $M_S(0,T)\sim (-\epsilon)^{\beta}$ and $\chi_0\sim \epsilon^{-\gamma}$ were identified through the examination of the obtained log scale $M(\epsilon)$ and $\chi(\epsilon)$ curves, respectively.
Subsequently, the magnetization curves, $M^{1/\beta}(\mu_0H/M)^{1/\gamma}$, were plotted using the derived values of exponents $\beta$ and $\gamma$.
The new values of $\beta$ and $\gamma$ were obtained from these magnetization curves.
This procedure was repeated until convergence of $\beta$, $\gamma$ and $T_C$ was achieved, which required 25 iterations.
The resulting values are of excellent precision, with the following values obtained: $\beta=0.337\pm0.001$, $\gamma=1.121\pm0.001$ and $T_C=146\pm1$~K.
The high degree of accuracy achieved is a consequence of the convergence of the completed procedure.
    
The inset of Fig.\ref{modArrott} illustrates the temperature dependence of the normalized slope (NS) of the $M^{1/\beta}(\mu_0H/M)^{1/\gamma}$ curves at $\mu_0H=9$~T.
The data illustrates that the obtained curves $M^{1/\beta}(\mu_0H/M)^{1/\gamma}$ are in fact nearly parallel to one another in the vicinity of $T_C$ at high fields.

\begin{figure}
\centering
\includegraphics[width=1.0\columnwidth]{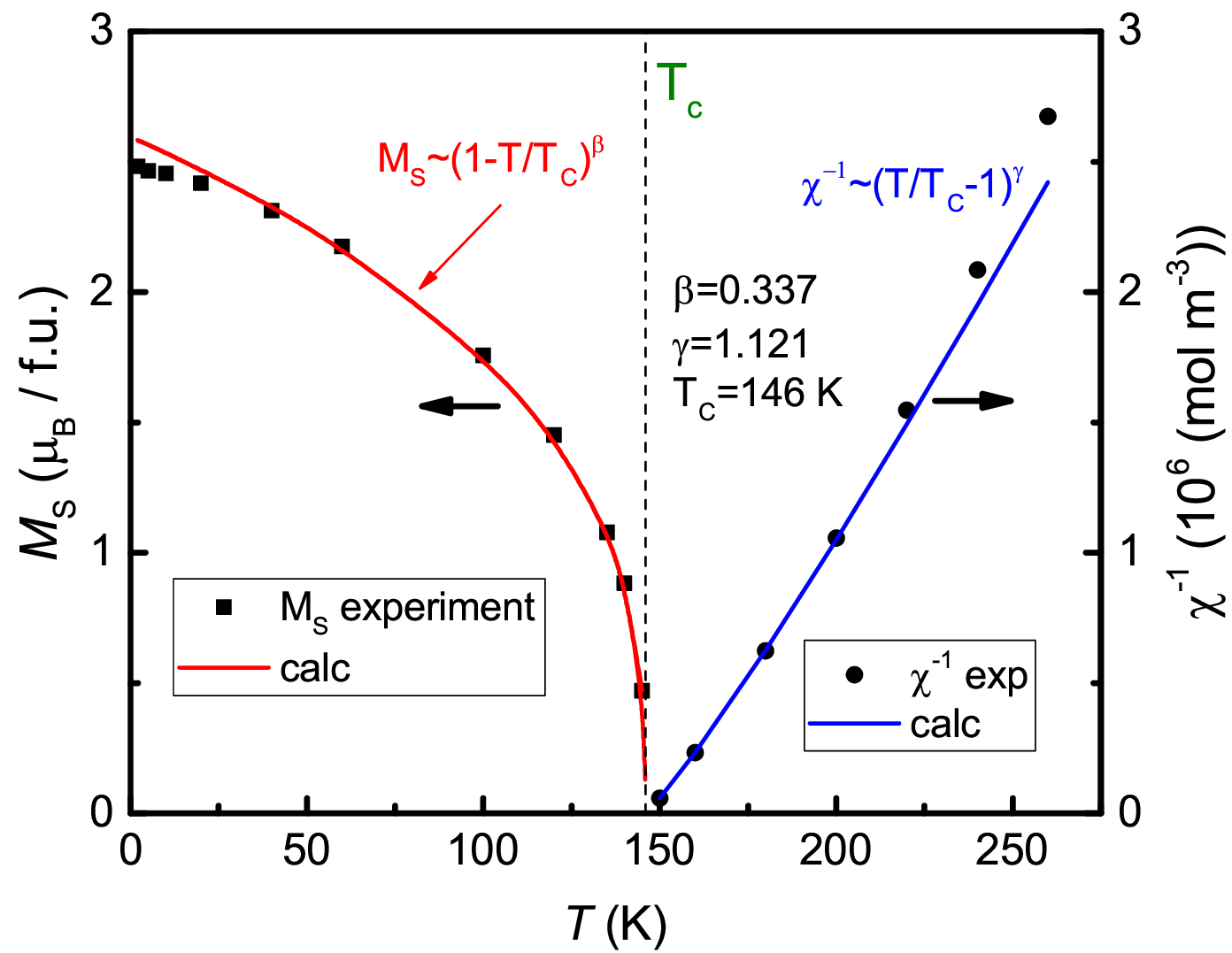}
\caption{The temperature dependence of spontaneous magnetization $M_S$ per formula unit (f.u.) (left axis, black dots) and inverse magnetic susceptibility $\chi^{-1}$ (right axis, black circles) obtained as the intercept of a tangent line to each $M^2(\mu_0H/M)$ curve (presented in the Fig.\ref{modArrott}) at $\mu_0H=9$~T with the ordinate or abscissa axes, respectively, for RhMnFeCoGe$_4$ compound. 
Red and blue lines represent curves calculated with equations \ref{eq_ms} and \ref{eq_hi} with $\beta=0.337$, $\gamma=1.121$ and $T_C= 146$~K.}
\label{crit_beh}
\end{figure}

The resulting $M_S(T)$ and $\chi^{-1}(T)$ experimental dependencies together with  curves calculated with equations \ref{eq_ms} and \ref{eq_hi} with $\beta=0.337$, $\gamma=1.12$ and $T_C= 146$~K are presented in the Fig.\ref{crit_beh}.

\begin{figure}
\centering
\includegraphics[width=1.0\columnwidth]{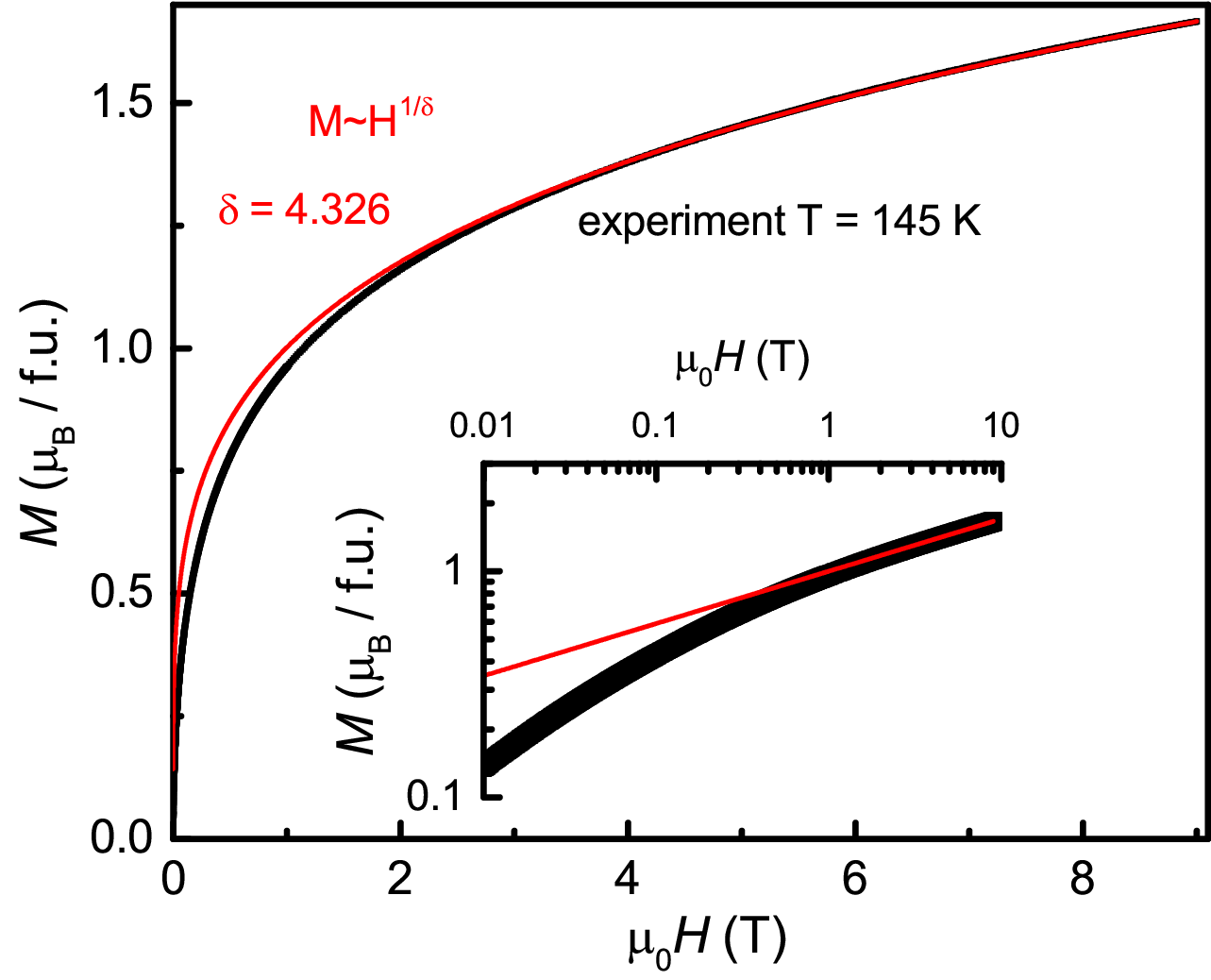}
\caption{The experimental curve for magnetic field dependence of magnetization $M$ per formula unit (f.u.)  at $T=145$~K  (black line) and the calculated curve, which was produced using the equation \ref{eq_delta} with the value of $\delta=4.326$ (red line). The inset is the same in log-log scale.}
\label{crit_M_H}
\end{figure}

The critical exponents $\beta$ and $\gamma$ are related to the third critical exponent $\delta$ through the Widom relation: $\delta=1+\gamma/\beta$.
Accordingly, the value of $\delta$ for the RhMnFeCoGe$_4$ compound can be determined to be $4.326\pm0.001$.
The exponent $\delta$ is responsible for determining the magnetization dependence with respect to magnetic field $\mu_0H$ at $T=T_C$ in sufficiently high magnetic fields:
\begin{equation}\label{eq_delta}
    M(H,T_C) \sim H^{1/\delta}, \, T=T_C.
\end{equation}
Figure \ref{crit_M_H} illustrates the magnetization curve $M(\mu_0H)$ of the RhMnFeCoGe$_4$ compound at $T=145$~K (black line).
The red line represents a calculated curve generated using the equation \ref{eq_delta} with $\delta=4.326$.
The inset depicts the same curves in log-log scale.
It can be seen that the equation \ref{eq_delta} accurately describes the experimental $M(\mu_0H)$ data at $T=T_C$ in high magnetic fields, as expected.
In light of these findings, it can be reasonably concluded that the Widom relation is an effective tool in this context.

The critical exponent $\beta=0.337$ in close proximity to the theoretical value of 0.325, as previously calculated for 3D-Ising model \cite{kaul1985static}.
The critical exponent $\gamma=1.12$ is situated at an intermediate point between the measured value of $\gamma$ for Gd (1.196) and the mean-field theory value of 1.00 \cite{kaul1985static}. 
Conversely, it can be proposed that the derived exponents are similar to those observed in the Ising model on the face-centered cubic lattice with $\beta= 0.312$ and $\gamma = 1.250$ \cite{muellner1974critical}.
It is well established that the critical exponents for the Ising ferromagnet with spin 1/2 and classical lattice gas are identical \cite{fisher1967thetheory}.
This implies that due to high disorder effects, RhMnFeCoGe$_4$ compound magnetically can be considered as having no preferred direction.

\subsection{NMR results}

\begin{figure}
\centering
\includegraphics[width=1.0\columnwidth]{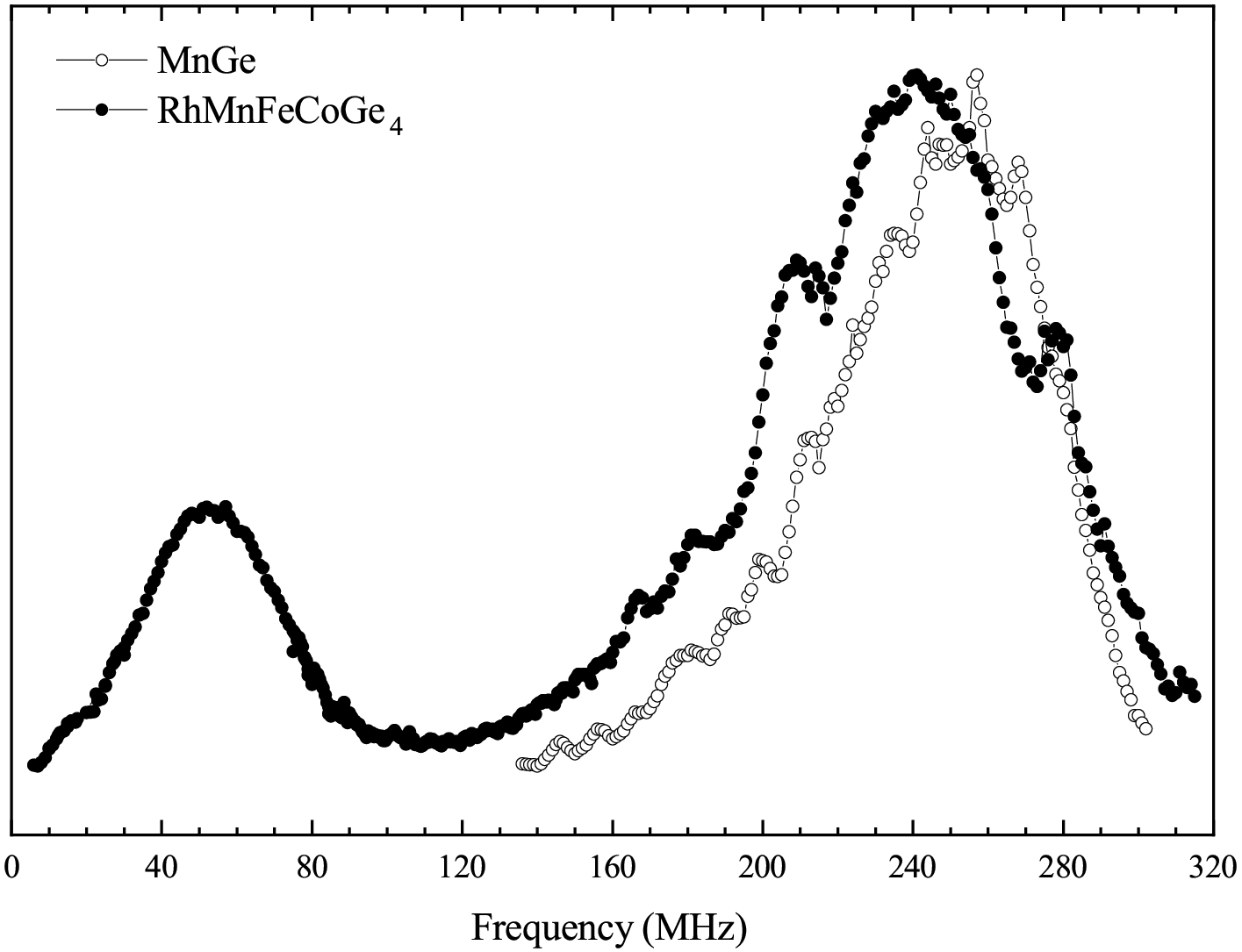}
\caption{NMR spectra for MnGe (open circles) and RhMnFeCoGe$_4$ (full circles) obtained in zero magnetic field at $T = 4.2$~K.}
\label{nmr}
\end{figure}

Figure \ref{nmr} shows the nuclear magnetic resonance (NMR) spectra for MnGe and RhMnFeCoGe$_4$, obtained in zero magnetic field at $T=4.2$~K. 
The NMR spectrum for MnGe demonstrates a single broad peak with the resonance frequency of $\nu_{res} = 257$~MHz, due to the signal from $^{55}$Mn nuclei. 
The spectrum of RhMnFeCoGe$_4$ exhibits two broad peaks: a low-frequency peak and a high-frequency peak with resonance frequencies 53~MHz and 241~MHz, respectively. 
As in MnGe, the high-frequency peak is associated with the $^{55}$Mn NMR signal in RhMnFeCoGe$_4$, whereas the low-frequency peak can be related to the signal from both $^{57}$Fe and $^{59}$Co nuclei. 
However, in unenriched samples, the signal from $^{57}$Fe nuclei is clearly discernible only in ferromagnets with substantial radio-frequency gain. 
Therefore, it is most likely that the low-frequency peak originates in the spectrum of RhMnFeCoGe$_4$ from the $^{59}$Co nuclei.

The magnetic field induced on manganese nuclei due to the hyperfine interaction of the nuclear spin with its nearest electron environment is the sum of two contributions \cite{allodi1997electronic}: $H_{loc} = A<\mu(Mn)> + \Sigma_j B_j <\mu(Mn)>$, where $A$ and $B$ are the constants of the hyperfine interaction. 
The first contribution to $H_{loc}$ is the main, as it caused by the interaction of manganese nucleus with its own electron shells. 
This term is proportional to the average magnetic moment on manganese atoms, $<\mu(Mn)>$. 
The second contribution is due to the effect of spin density transfer from neighboring magnetic ions, and it is about an order smaller than the first term.

On the other hand, the $H_{loc}$ value is determined experimentally from the relation  \cite{slichter2013principles}: $H_{loc} = \nu_{res}/{}^{55}\gamma$, where $\nu_{res}$ is the resonance frequency and ${}^{55}\gamma$ is the gyromagnetic ratio of manganese nuclei. 
It is known from Ref.\cite{makarova2012neutron} that for parent compound MnGe the magnetic moment $<\mu(Mn)>=2.3$~$\mu_B$ at $T = 2$~K and therefore the constant $A$ can be estimated as $A = -106$~kOe/$\mu_B$.
This value differs from the previously obtained $A = - 138$~kOe/$\mu_B$ for MnSi \cite{yasuoka1978nmr}, but is very close to $A = - 110$~kOe/$\mu_B$ from Ref.\cite{ghorai2009synthesis}.

It is reasonable to assume that the hyperfine interaction constants do not differ significantly for MnGe and RhMnFeCoGe$_4$. 
Consequently, the magnetic moment on manganese in RhMnFeCoGe$_4$ estimated from the ${}^{55}$Mn NMR spectrum (Fig. \ref{nmr}) is $\approx 2.2\, \mu_B$.
Furthermore, assuming the value of the constant $A$ for one 3\emph{d} electron is about $-100$~kOe/$\mu_B$  \cite{freeman1965hyperfine}, the magnetic moment of cobalt can be determined $\sim0.5$~$\mu_B$. 

\subsection{Resistivity}

\begin{figure}
\centering
\includegraphics[width=1.0\columnwidth]{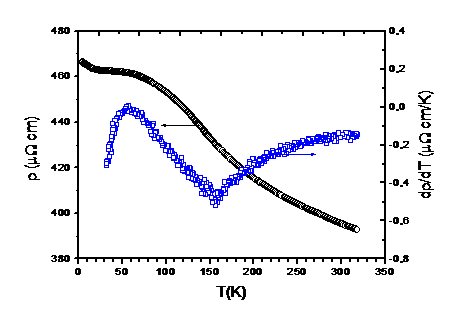}
\caption{The temperature dependence of AC resistivity and its temperature derivative of the RhMnFeCoGe$_4$ compound.}
\label{ro}
\end{figure}

AC resistivity of RhMnFeCoGe$_4$ compound demonstrates a slight increase with decreasing temperature (Figure \ref{ro}), a characteristic observed in semimetals.
Nevertheless, in all binary compounds (RhGe, MnGe, FeGe, and CoGe), there is a notable reduction in electrical resistance with decreasing temperature\cite{pedrazzini2007metallic, deutsch2014twostep, tsvyashchenko2016superconductivity, baek2022possible}. 
As illustrated in Figure \ref{ro}, the temperature dependence of the derived resistivity exhibits a sharp transition at $T_C = 150$ K, which corresponds to the magnetic transition.

\subsection{Effect of high pressure on critical temperature}

\begin{figure*}
\centering
\includegraphics[width=1.0\columnwidth]{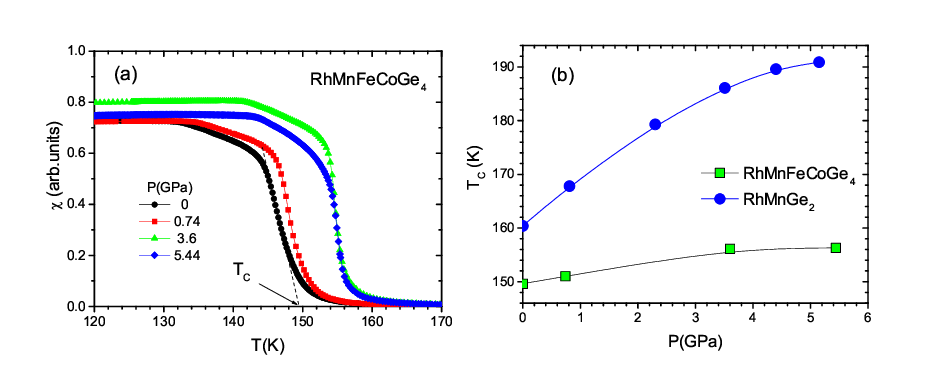}
\caption{The temperature dependence of the magnetic susceptibility of RhMnFeCoGe$_4$ near $T_C$ at varying pressures (panel A). Panel B depicts the magnetic $P-T$ diagrams of RhMnFeCoGe$_4$ (this work) and RhMnGe$_2$ (ref. \cite{sidorov2018magnetic}).}
\label{press}
\end{figure*}

The impact of elevated pressure on the magnetic transition temperature ($T_\mathbf{C}$) of RhMnFeCoGe$_4$ was investigated through the utilisation of magnetic susceptibility measurements at varying pressures. 
The result is presented in Fig.\ref{press}A. 
The temperature at which the magnetic susceptibility exhibited a sharp rise was defined as $T_C$.
The resulting magnetic $P-T$ diagram is depicted in Fig.\ref{press}B. 
Initially, $T_C$ increases with the increase of pressure at a rate $dT_{C}/dP = 1.9$~K/GPa.
However, above 3.6~GPa, the increase of $T_C$ terminates. 
For comparison, the magnetic $P-T$ diagram of the ternary compound RhMnGe$_2$ (\cite{sidorov2018magnetic}) is shown in Fig.\ref{press}B as well.

It is noteworthy that for binary compounds FeGe and MnGe, the magnetic transition temperature ($T_C$) is observed to decrease at elevated pressures. 
Furthermore, at sufficiently high pressures, the magnetism in these compounds is completely suppressed \cite{pedrazzini2007metallic,deutsch2014twostep,martin2016magnetovolume}.
The initial rate of decrease of $T_C$ at high pressure for MnGe is $dT_{C}/dP = - 6.9$~K/GPa \cite{sidorov2018magnetic}.
The substitution of Rh atoms for 30\% of Mn atoms results in a positive value for $dT_{C}/dP$, which is equal to $4$~K/GPa.
When  Rh atoms are substituted for 50\% of Mn atoms (in the case of RhMnGe$_2$) the value of $dT_{C}/dP$ increases to 9.1~K/GPa \cite{sidorov2018magnetic}.
Figure \ref{press}B illustrates the impact of elevated pressure and the incorporation of Fe and Co atoms on the pressure dependence of $T_C$.
Initially, it is evident that for RhMnGe$_2$, $dT_{C}/dP$ exhibits a gradual decline at elevated pressure, with a saturation value near 6~GPa.
A similar trend  is observed for RhMnFeCoGe$_4$, even at lower pressures.
Furthermore, the addition of Fe and Co results in a nearly fivefold decrease in the initial value of $dT_{C}/dP$ for RhMnFeCoGe$_4$ in comparison to RhMnGe$_2$.
This is attributed to the negative value of $dT_{C}/dP$ in FeGe \cite{pedrazzini2007metallic} and the absence of long-range magnetism in CoGe \cite{baek2022possible}.
The substitution of an Rh atom for 10 at.\% of the Fe atoms in FeGe (\cite{salamatin2022thehyperfine}) results in an initial rate of $dT_{C}/dP = 1.25$~K/GPa, which is positive. 
However, at 3~GPa $T_C$ reaches its maximum, and at higher pressures $T_C$ decreases rapidly, as observed in FeGe. 
A review of Fig.\ref{press}B suggests that the $T_C$ of RhMnGe$_2$ and RhMnFeCoGe$_4$ may begin to decline above 6~GPa, in a manner similar to that observed in MnGe, FeGe, and Fe$_{0.9}$Rh$_{0.1}$Ge.

\subsection{Ab initio \emph{calculations}}

The real high-pressure synthesized material was simulated as an idealized ordered compound RhMnFeCoGe$_4$ with the same stoichiometry.
After relaxation, the B20 lattice constant in the FM state is $a = 4.7539$~\AA, thus, the theoretical specific volume is only about 1\% less than the experimental one. 
The paramagnetic (PM) state is less favorable in energy than the FM state, and its specific volume is 1.5\% less than that of the FM state.

Our calculated magnetic moment per formula unit is equal to 3.2 $\mu_B$, which exceeds the experimental value of 2.5 $\mu_B$. 
This discrepancy may be ascribed to chemical disorder, residual microstresses, and other defects which can exist in high-pressure phases. 
Thus, in experiments, an average moment was measured, while in simulations, the magnetic moment was evaluated within a simple model of collinear arrangement of spins.

The theoretical moment is mainly contributed by the Mn and Fe atoms: $m_{Mn} = 2.0$ and $m_{Fe} = 1.3$ $\mu_B$. 
The Co moment is noticeably lower, while the induced moments on Rh (positive) and Ge (negative) are markedly insignificant in magnitude (for details, please refer to the first row of Table \ref{t_calc}). 
The theoretical Mn moment is close to the value of 2.2~$\mu_B$ estimated from our NMR experiments, while the calculated Co moment of 0.16~$\mu_B$ is smaller, but not much, than the NMR-estimated value of 0.5~$\mu_B$.
This discrepancy may be attributed to the presence of supplementary magnetic fields at the Co sites, which are induced by neighboring Mn and Fe atoms with high magnetic moments. 
This resulted in an increased measured Co moment in the experiment in comparison with that of the "bare" calculated Co moment.

\begin{table*}[t]
\centering
\caption{The results of WIEN2k calculations: Magnetic moments $m$ [$\mu_B$], hyperfine magnetic fields $H_{hf}$ [kOe], and the pairing constants $A = H_{hf}/\mu_{TM}$ [kOe/$\mu_B$]. The 1$^{st}$ row lists the magnetic moments calculated using VASP.}
\begin{tabular}{|c|c|c|c|c|c|c|}
\hline
  & Rh (\AA) & Mn & Fe    & Co & Ge & f.u.\\
\hline
 $m_{VASP}$ & 0.02 & 2.00 & 1.32  & 0.16 & $<-0.17>$ & 3.22\\
\hline
 $m_{WIEN2k}$ & 0.02 & 1.92 & 1.27  & 0.15 & $<-0.05>$ & 3.05\\
\hline
 $H_{hf}$ & -112.9 & -135.1 & -118.1  & -70.6 & $<-164.0>$ & -\\
\hline
 A & - & -71.4 & -93.0  & -470.7 & - & -\\
\hline
\end{tabular}
\label{t_calc}
\end{table*}

Table \ref{t_calc} additionally presents the magnetic and hyperfine characteristics calculated using an alternative approach, namely the all-electron LAPW+lo method as implemented in the WIEN2k package \cite{blaha2020wien2k}.
This all-electron approach is quite suitable for exploring the immediate vicinity of a nucleus, where hyperfine interactions are significant. 
We briefly list the calculation parameters. 
The MT-radii of Rh, Mn, Fe, Co, and Ge were set to 2.25, 2.21, 2.18, 2.17, and 2.06 Bohr, respectively. 
The energy cutoff parameter $R_{min}K_{max}$ was equal to 8.0 and the maximal number of k points was 451.
The lattice parameters used in this calculation were already relaxed by VASP, so many steps of additional WIEN2k relaxation were not required.
Table \ref{t_calc} shows average values for Ge, since the positions and, correspondingly, the physical parameters, of the four Ge atoms are slightly different.
The magnetic moments found with WIEN2k are close to, but slightly smaller than, those obtained with VASP.
The reason lies in the geometry of the LAPW basis, which implies the presence of an interstitial (IS) part of the cell. 
In this case, the interstitial moment $m_{IS} = -0.14$ $\mu_B$.

\begin{figure*}
	\centering
	\includegraphics[width=0.4\textwidth]{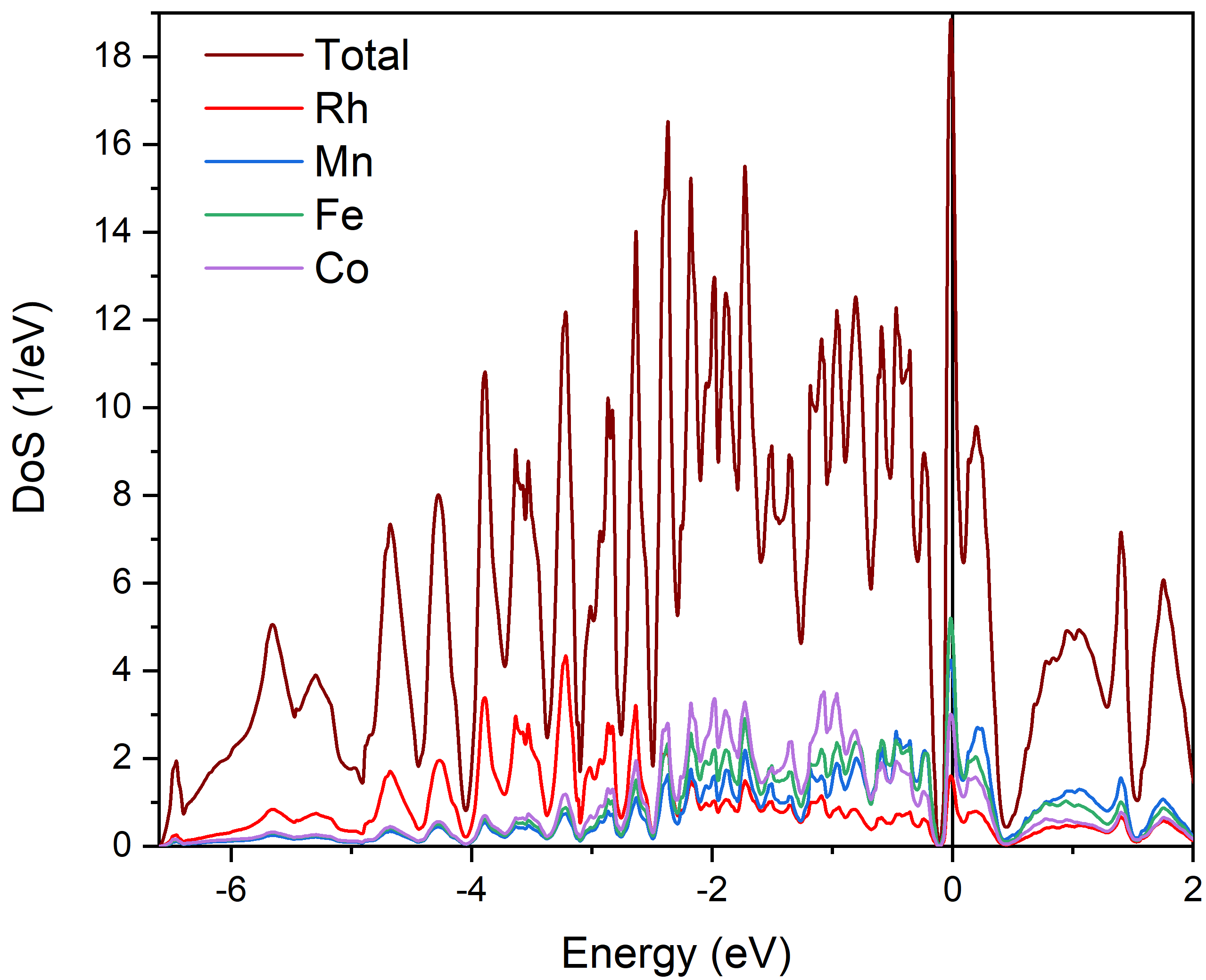}\hspace{5pt}
    \includegraphics[width=0.4\textwidth]{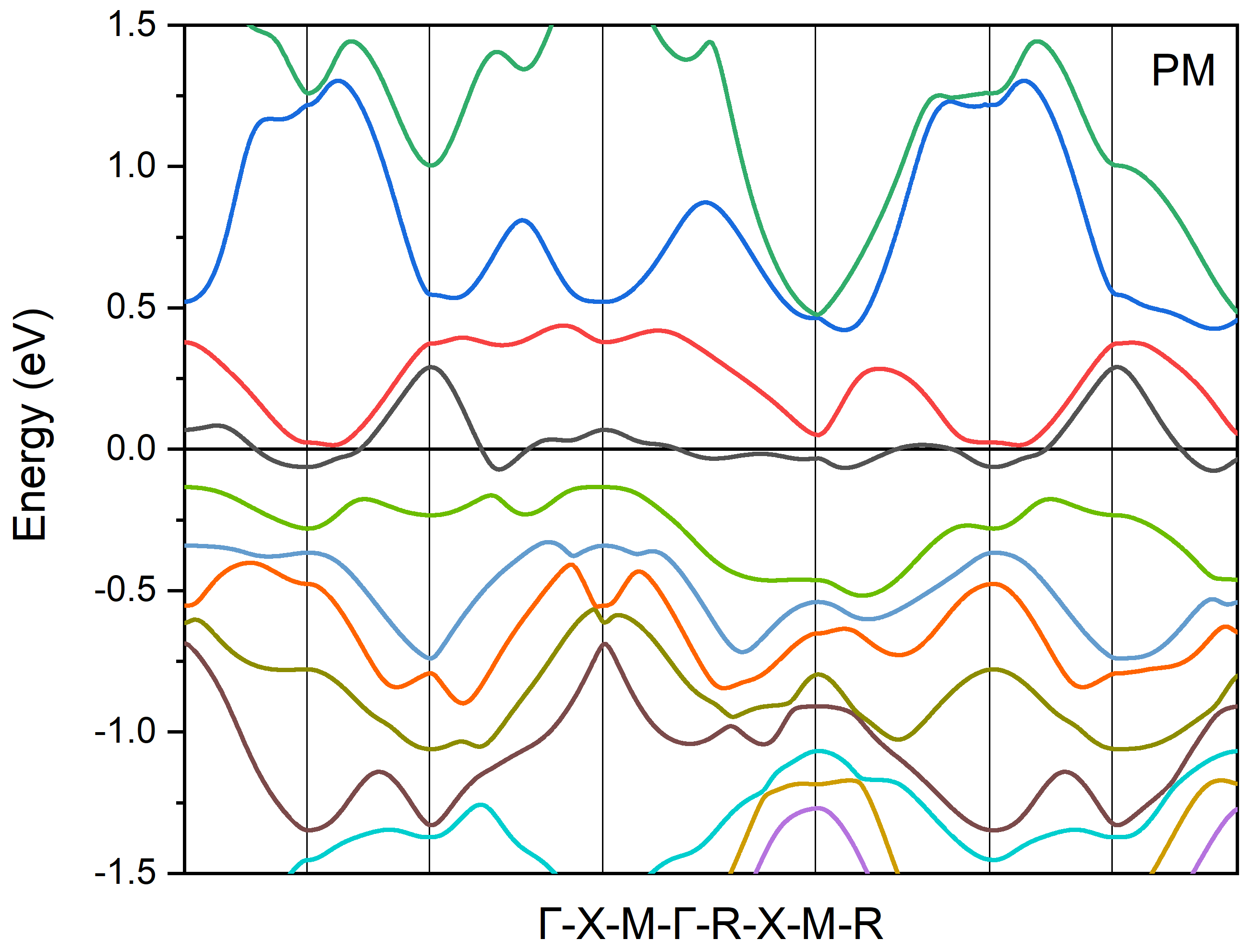}
	\caption {
    PM RhMnFeCoGe$_4$. The total and atom-projected density of states (left) and the band structure (right). A small contribution of Ge states to DOS is not shown. Energy is measured from the Fermi level.
    }
	\label{fig_calc_pm}
\end{figure*}

\begin{figure*}
	\centering
	\includegraphics[width=0.4\textwidth]{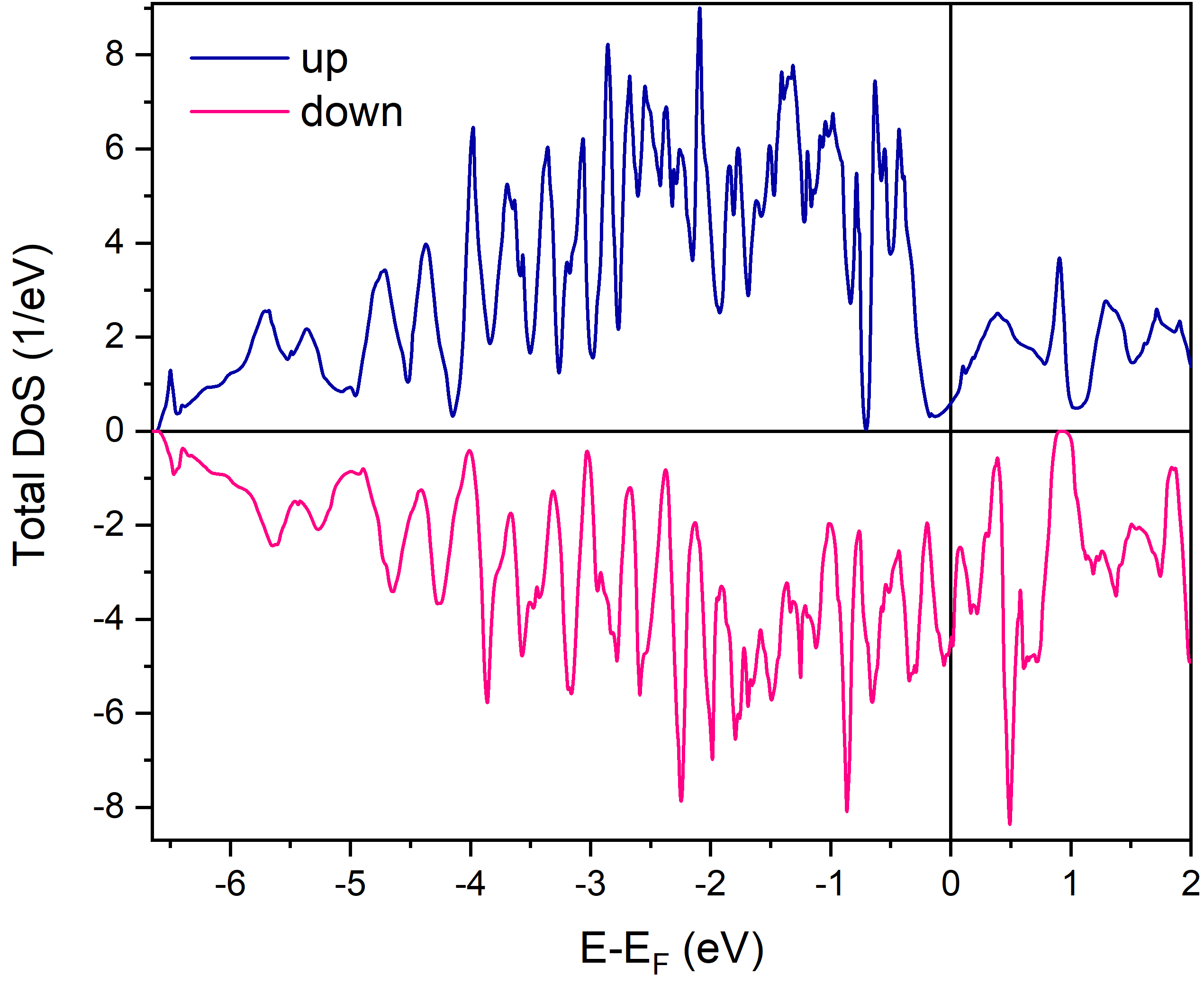}\hspace{5pt}
    \includegraphics[width=0.4\textwidth]{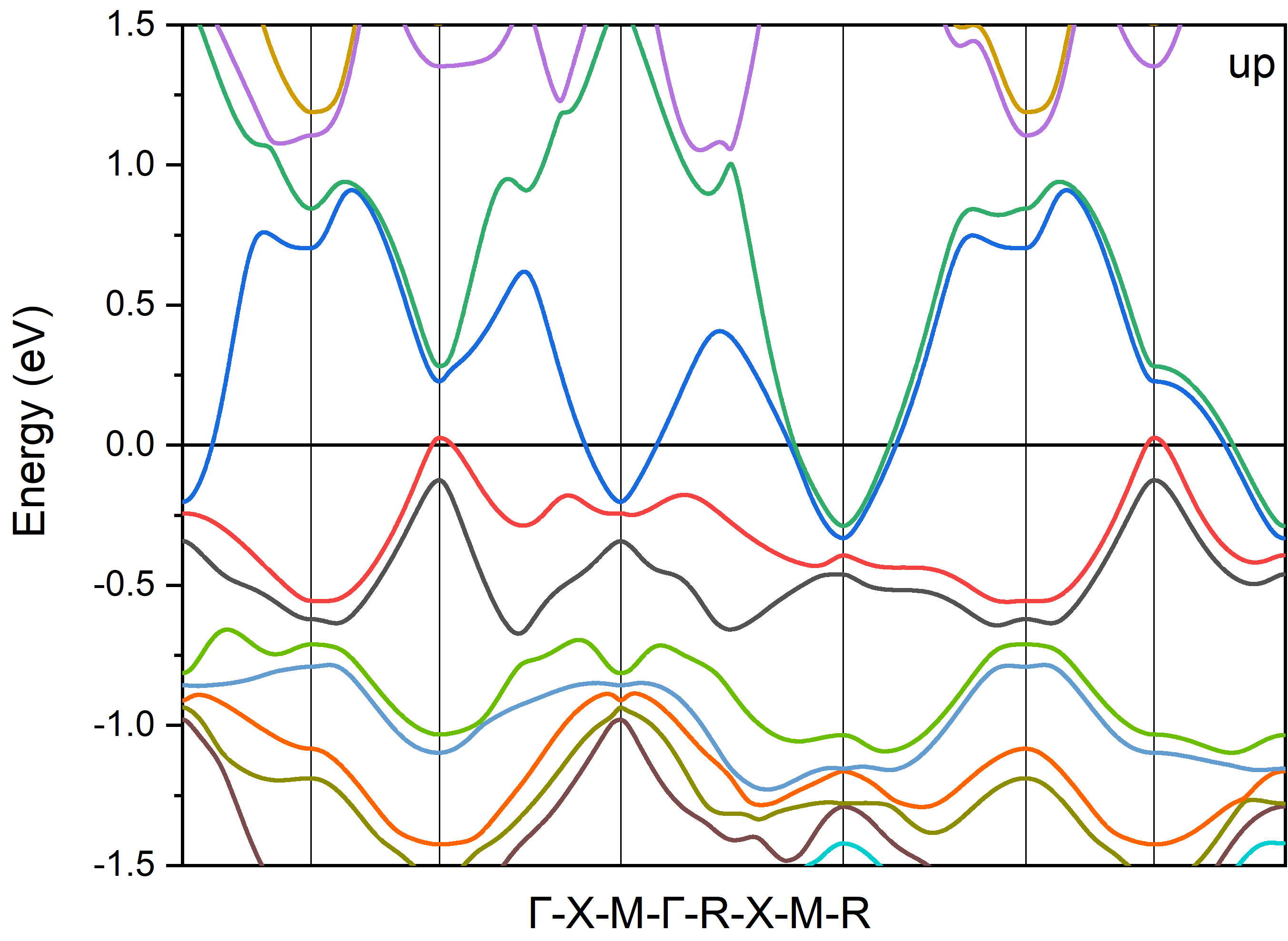}\vspace{5pt}
    \includegraphics[width=0.4\textwidth]{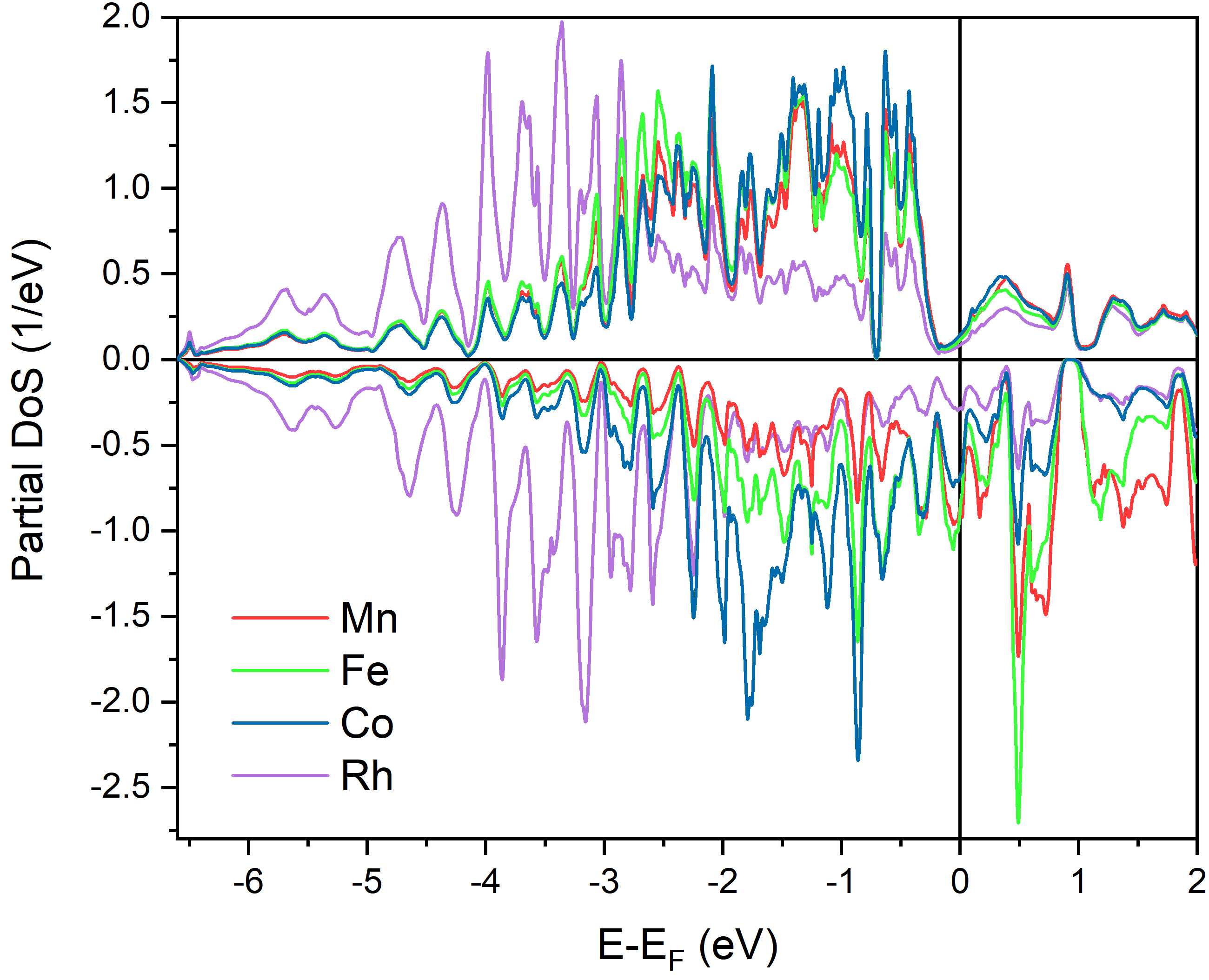}\hspace{5pt}
    \includegraphics[width=0.4\textwidth]{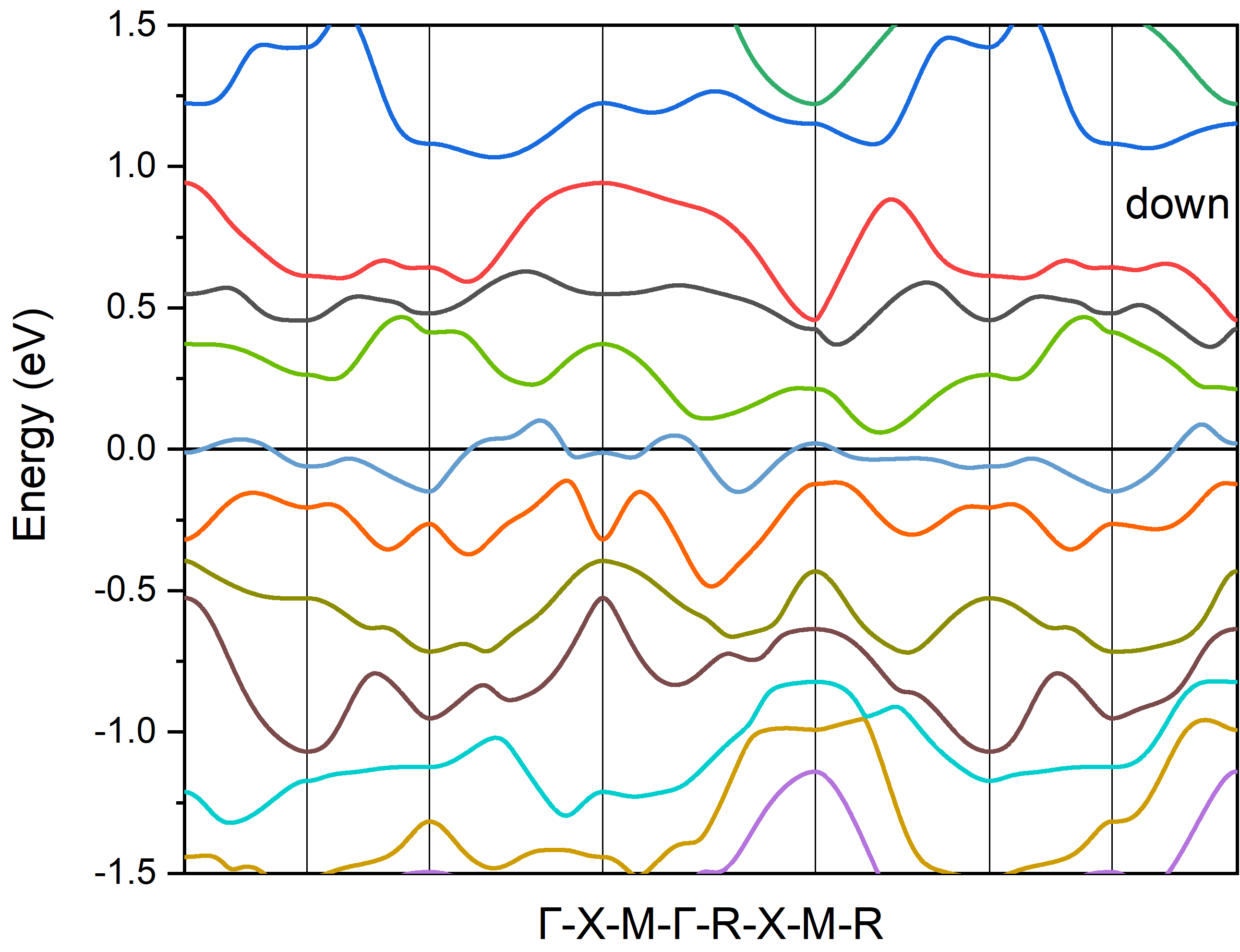}
	\textbf{(b)}
	\caption {
    FM RhMnFeCoGe$_4$. Left: the total (top) and partial (bottom) density of states. A small contribution of Ge states is not shown. Right: the band structure for up (top) and down (bottom) spin directions. Energy is measured from the Fermi level.
    }
	\label{fig_calc_fm}
\end{figure*}

Figures \ref{fig_calc_pm} and \ref{fig_calc_fm} show the density of states (DOS) and the band structure for paramagnetic (PM) and FM RhMnFeCoGe$_4$, respectively.

One can see a high value of the DOS at the Fermi level, $N(E_F)$, in the PM state and a moderate value of $N(E_F)$ in the FM state because of spin splitting in magnetic case.
The behavior of DOS near the Fermi energy corresponds to the metallic type of conductivity. 
Thus, an increase in resistivity with decreasing temperature found in our measurements, typical of semimetals, is most likely due to chemical disorder in the real material, which is not taken into account in our simulation.
It is evident from the figures that the shape of bands for RhMnFeCoGe$_4$ has traces of symmetry features characteristic of B20 binary compounds RhGe, MnGe, FeGe, and CoGe. 

\section{Conclusions}

A high-entropy compound, RhMnFeCoGe$_4$, was synthesized in the cubic B20 structure under conditions of high pressure and high temperature.
This compound exhibits ferromagnetic order at a critical temperature, $T_C=146$~K.
The ordered moment is 2.5 $\mu_B$ per f.u. at $T = 2$~K.
No discernible hysteresis is evident.
In the vicinity of the critical temperature, the critical exponents are $\beta=0.337$ and $\gamma=1.121$. 
These values are in close proximity to those predicted by the theoretical model of the Ising ferromagnet.
NMR spectra for MnGe and RhMnFeCoGe$_4$ compound are obtained at $T=4$~K.
The magnetic moments of Mn and Co were determined from NMR spectra to be 2.2~$\mu_B$ and 0.5~$\mu_B$, respectively.
The obtained moments are in reasonable agreement with those obtained from \emph{ab initio} calculations. 
The density of states and band structure are obtained for both the paramagnetic and ferromagnetic states.
The critical temperature is increasing in the RhMnFeCoGe$_4$ compound under the influence of compression.

\section{Acknowledgments}
This research was funded by the Russian Science Foundation Grant No. 25-12-68013 (22-12-00008-$\pi$).



 \bibliographystyle{elsarticle-num} 
 \bibliography{cas-refs}





\end{document}